\documentclass[aps,pre,preprint,groupedaddress,showpacs]{revtex4-1}
\pdfoutput=1

\usepackage{amsmath,amssymb}
\usepackage{graphicx,color}
\usepackage{mathrsfs}
\usepackage{hyperref}

\begin{document}

\title{Master equation for She-Leveque scaling and its classification in terms of other Markov models of developed turbulence}
\author{Daniel Nickelsen}
\email[]{danielnickelsen@sun.ac.za}
\affiliation{National Institute for Theoretical Physics (NITheP), Stellenbosch, South Africa}
\affiliation{Institute of Theoretical Physics, University of Stellenbosch, South Africa}
\affiliation{Institute for Physics, Carl-von-Ossietzky University, Oldenburg, Germany}

\date{\today}

\begin{abstract}
The statistics of velocity increments in homogeneous and isotropic turbulence exhibit universal features in the limit of infinite Reynolds numbers. After Kolmogorov's scaling law from 1941, many turbulence models aim for capturing these universal features, some are known to have an equivalent formulation in terms of Markov processes. We derive the Markov process equivalent to the particularly successful scaling law postulated by She and Leveque. The Markov process is a jump process for velocity increments $u(r)$ in scale $r$ in which the jumps occur randomly but with deterministic width in $u$. From its master equation we establish a prescription to simulate the She-Leveque process and compare it with Kolmogorov scaling. To put the She-Leveque process into the context of other established turbulence models on the Markov level, we derive a diffusion process for $u(r)$ using two properties of the Navier-Stokes equation. This diffusion process already includes Kolmogorov scaling, extended self-similarity and a class of random cascade models. The fluctuation theorem of this Markov process implies a ``second law'' that puts a loose bound on the multipliers of the random cascade models. This bound explicitly allows for instances of inverse cascades, which are necessary to satisfy the fluctuation theorem. By adding a jump process to the diffusion process, we go beyond Kolmogorov scaling and formulate the most general scaling law for the class of Markov processes having both diffusion and jump parts. This Markov scaling law includes She-Leveque scaling and a scaling law derived by Yakhot.
\end{abstract}

\maketitle

\newcommand{\lla}{\left\langle}
\newcommand{\rra}{\right\rangle}
\newcommand{\pt}{\partial}
\newcommand{\laplace}{\Delta}
\newcommand{\vrms}{v_\mr{rms}} 
\newcommand{\lch}{\ell_\mr{ch}} 
\newcommand{\vch}{v_\mr{ch}} 
\newcommand{\dfr}{d_\mr{fr}} 
\newcommand{\Df}{F}
\newcommand{\Dg}{D}
\newcommand{\tb}{c}
\newcommand{\tB}{\bar B_k}
\newcommand{\dk}{d_k}
\newcommand{\DS}{\Delta S} 
\newcommand{\Ds}{\Delta s} 
\newcommand{\Dr}{\Delta r} 
\newcommand{\tu}{\tilde u} 
\newcommand{\jm}{\chi} 
\newcommand{\jd}{\theta} 
\newcommand{\Ak}{\varPsi^{^{\text{\tiny ($k$)\!\!}}}} 
\newcommand{\Af}{\varPsi^{^{\text{\tiny ($1$)\!\!}}}}
\newcommand{\Ag}{\varPsi^{^{\text{\tiny ($2$)\!\!}}}}
\newcommand{\dd}{\mathrm{d}} 
\newcommand{\mr}[1]{\mathrm{#1}}
\newcommand{\vek}[1]{\boldsymbol{#1}} 
\newcommand{\Rey}{\mr{R\hspace{-0.9pt}e}} 
\newcommand{\eps}{\varepsilon} 
\newcommand{\beps}{\bar\eps}
\newcommand{\e}{\epsilon}
\newcommand{\z}{\zeta}
\newcommand{\OO}{\mathcal{O}}

\section{Introduction}\label{sec:intro}
A turbulent flow is a particularly challenging example of a complex system. One reason for that is that with increasing turbulence intensity, the number of effective degrees of freedom in a turbulent flow increases until literally each fluid molecule with their position and momentum have to be considered \cite{Kraichnan1958,landau1987fluid}. Furthermore, the dynamics is chaotic and depends sensitively on initial conditions. This is a situation comparable to problems considered in statistical mechanics, suggesting a statistical analysis of turbulence.

In 1941, Kolmogorov was the first to apply a statistical analysis to homogeneous and isotropic turbulence \cite{Kolmogorov1941a,Kolmogorov1941b,Kolmogorov1941c}. Central to this analysis was the prediction of universal scaling laws. Finding scaling exponents that universally reflect the velocity increment statistics observed in experiments and simulations has been the objective of turbulence research ever since. Although successful scaling laws have been found \cite{Kolmogorov62JoFM,SheLeveque94PRL,Yakhot98PRE}, they still evade a profound understanding of the underlying mechanism \cite{Frisch95}. In particular the phenomenon of small-scale intermittency, where strong fluctuations on small scales seem to appear out of nothing, remains to be a focus of current research \cite{Sreenivasan1997,Nickelsen2013,Faranda2014,Biferale2014,Ali2016,Friedrich2016}.

The most important mechanism responsible for the emergence of scaling laws stems from the self-similarity found in turbulent flows, where structures on large scales, e.g. vortices, repeat themselves on smaller scales \cite{Richardson22}. This mechanism is generally assumed to be the turbulent cascade, in which turbulent structures become unstable and break into similar structures.

The picture of the turbulent cascade has not only inspired researchers to analyse the scaling symmetry of turbulent flows. It is also at the core of a Markov analysis based on perceiving the turbulent cascade as realisations of a Markov process, which was introduced by Friedrich and Peinke \cite{FriedrichPeinke97PRL}. In this analysis, the drift and diffusion coefficients of a Fokker-Planck equation are determined directly from experimental data. The key observation in \cite{FriedrichPeinke97PRL} is that turbulent cascades exhibit the Markov property which allows to reproduce the complete multi-scale statistics of velocity increments from the Fokker-Planck equation. The Markov analysis by Friedrich and Peinke has been improved substantially and led to many insights in the last decades \cite{Renner2001,NawroPeinkKleinFried07PRE, GottschallPeinke08NJoP, Renner2011, Kleinhans2012}. Apart from developed turbulence, the Markov analysis has become popular for other settings of turbulence, such as turbulent boundary layers \cite{Tutkun2017} or wind energy \cite{Milan2013}. An overview of applications of the Markov analysis to many complex systems can be found in \cite{Friedrich2011a}. Other Markov approaches on a more microscopic level have been pioneered by Kraichnan \cite{Kraichnan1971} and Frisch and collaborators \cite{Frisch1974} in which stochastic terms are introduced to the Navier-Stokes equation. A more direct approach is to use the Markov property to construct Lagrangian trajectories and map the resulting statistics of flow velocity to Eulerian coordinates \cite{Pedrizzetti1994}. In the present work we are concerned with stochastic approaches on a more descriptive level which exploit the Markov character of the turbulent cascade instead of explicitly employing the Navier-Stokes equation.

The Markov property as an approach to complexity was already used in Einstein's theory of Brownian motion \cite{Einstein1905}, in which the outcome of collisions is rendered as a random process instead of resolving exact trajectories of Brownian particles. The setting of this particular Markov process enables a thermodynamic interpretation: During collisions, heat is transferred between the Brownian particles and the medium. Building on this perception, stochastic expressions for heat and other thermodynamic quantities like work and entropy were defined \cite{Sekimoto1998}, leading to a complete thermodynamic picture of driven Brownian motion. Colloidal particles is only one example of systems studied in the field now known as stochastic thermodynamics, see \cite{Seifert2012} and references therein for an introduction to the field. One celebrated result of stochastic thermodynamics are fluctuation theorems which tighten the second law to equalities. Coming back to turbulent cascades, in our recent work \cite{Nickelsen2013} we made use of the fact that Markov processes are central in both the Markov analysis of turbulence and in stochastic thermodynamics to show that a fluctuation theorem also holds for the turbulent cascade and constitutes a measure for the correct modelling of small-scale intermittency. 

In this paper, we expand further on the role Markov processes play in developed turbulence and demonstrate that they arise naturally from the phenomenology of the turbulent cascade. Some Markov processes are already known to be equivalent to established turbulence models, such as Kolmogorov scaling \cite{FriedrichPeinke97PRL}, log-normal random cascade models \cite{Amblard1999} and Yakhot's approach to turbulence \cite{Davoudi1999}. These findings are scattered in the literature and deserve to be put together in order to reveal their systematics in the Markov description. We review these models together with their equivalent Markov processes and demonstrate how the picture of the turbulent cascade leads to a simple diffusion process which already represents a number of turbulence models including extended-self similarity as a special case of log-normal random cascade models. The fluctuation theorem of this diffusion process implies a second law for the turbulent cascade which we find to put a loose bound on the multipliers in random cascade models. We then extent the list of Markov processes representing turbulence models by including jump processes as part of the Markov process. Formulating the most general scaling law for the class of Markov processes having both diffusion and jump parts enables us to find the Markov processes that reproduce scaling laws postulated by She and Leveque \cite{SheLeveque94PRL} and derived by Yakhot \cite{Yakhot98PRE}. In particular, being our main result, we formulate the Master equation (\ref{eq:SL_PDE}) for the popular She-Leveque scaling. We conclude with table \ref{tabres} compiling all considered Markov processes that represent turbulence models, demonstrating which components of the Markov processes distinguish the various models.

\section{Approaches to developed turbulence}\label{sec:models}
We begin with the traditional approach to capture universal features of developed turbulence and give a survey on various established turbulence models.

The ruling equation of turbulent flows is the Navier-Stokes equation which reads in dimensionless quantities \cite{landau1987fluid,Frisch95},
\begin{equation} \label{eq:NSE_dimless}
 	\frac{\pt\vek{v}(\vek{x},t)}{\pt t} + \left(\vek{v}(\vek{x},t)\cdot\vek{\nabla}\right)\vek{v}(\vek{x},t) = \frac{1}{\Rey}\laplace\vek{v}(\vek{x},t) -\vek{\nabla}P(\vek{x}) + \vek{f}(\vek{x},t) \,.
\end{equation}
Here, $\vek{v}(\vek{x},t)$ is the flow velocity field for position $\vek{x}$ and time $t$, $P(\vek{x})$ is the pressure field, and $\vek{f}(\vek{x},t)$ is the external forcing that generates turbulence. The Reynolds number $\Rey$ relates forces of turbulence generation with viscous forces by $\Rey=\frac{\lch \vch}{\nu}$, where $\lch$ and $\vch$ are the characteristic length and velocity of turbulence generation and $\nu$ is the kinematic viscosity. As the dissipation term $\laplace\vek{v}(\vek{x},t)$ enters with the prefactor $1/\Rey$, large Reynolds numbers indicate a subordinate role of dissipation. In this paper we are mainly concerned with the limit of infinite Reynolds numbers.

The Navier-Stokes equation is too complex to be solved analytically. However, we will exploit this complexity by building upon two known properties \cite{landau1987fluid,Frisch95}:
\begin{itemize}
	\item[(a)] The dynamics of the Navier-Stokes equation is chaotic and sensitively depends on the initial conditions.
	\item[(b)] Turbulent structures are unstable under the non-linear dynamics of the Navier-Stokes equation and break-up into smaller structures.
\end{itemize}

The important consequence of (a) is that a turbulent flow evolving freely after its generation (e.g. after a grid) does not bear any resemblance with the structure of the generation. The turbulent flow is hence ruled by the force-free Navier-Stokes equation with a non-equilibrium initial condition which encodes a set of turbulent structures of large scale $L<\lch$. Due to (b), these structures break up and form the \textit{turbulent cascade} in which turbulent structures transport energy from large to small scales with an average rate of $\beps$. If furthermore the Reynolds number is sufficiently large, effects of dissipation are negligible for scales larger than the Taylor length scale $\lambda$. The range of scales between $\lambda$ and $L$ is known as \textit{inertial range} in which neither turbulence generation nor dissipation play a role \cite{Frisch95}.

The force-free Navier-Stokes equation has symmetries which entail properties of the flow field $\vek{v}(\vek{x})$, among which are homogeneity, isotropy, and, for $\Rey\to\infty$, scaling symmetries \cite{Frisch95}. A homogeneous and isotropic flow field are the defining properties of \textit{developed turbulence}. The scaling symmetry of developed turbulence corresponds to the self-similarity of a turbulent flow and is a focus of turbulence research. Self-similarity expresses itself as scaling laws in a statistical analysis of the structures in the flow field $\vek{v}(\vek{x},t)$, as we discuss now.

We assume the turbulent flow to be fully developed and probe structures in the flow by the velocity increments
\begin{equation} \label{eq:vel_incr_full}
 u_{\vek{x},\vek{e},t}(\vek{r}) = \vek{e}\vek{v}(\vek{x} + \vek{r},t) - \vek{e}\vek{v}(\vek{x},t)
\end{equation}
projected on the unit vector $\vek{e}$. As the chaotic property (a) imposes a certain randomness in the flow field, we understand $\vek{v}(\vek{x},t)$ to be a correlated random field with correlation length $L$. By fixing $\vek{e}$ and $t$ and changing $\vek{x}$ in steps sufficiently larger than $L$, application of the above definition yields a set of realisations $u(\vek{r})$ of an underlying stochastic process. Owing to a homogeneous and isotropic flow field, the realisations $u(\vek{r})$ are independent and follow the same probability density function $p(u,\vek{r})$ at fixed $r$.

As developed turbulence is by definition independent from its generation, the statistics of $u(\vek{r})$ can only arise from the internal non-linear dynamics of the force-free Navier-Stokes equation for scales sufficiently smaller than the scale of turbulence generation. Therefore, the statistics of $u(\vek{r})$ is assumed to be universal in the inertial range and the turbulent cascade is considered to be the fundamental mechanism defining the statistics. In the following, we will be concerned with these universal features of developed turbulence.

To simplify, we set $\vek{e}=\vek{e}_x$, fix a certain $t$, and consider the one-dimensional longitudinal velocity increments
\begin{equation} \label{eq:vel_incr}
 u(r,x) = v(x + r) - v(x) \,,
\end{equation}
where $v$, $x$ and $r$ now denote the $x$-component of the previous vectors. It is common to analyse their statistics by the moments
\begin{equation} \label{eq:FDT_moments}
 S^n(r) = \lla u(r)^n \rra = \int u(r) \, p(u,r) \, \dd u
\end{equation}
which, as they are used to probe the structures of the flow, are known as structure functions of $n$-th order. Since $u(r)$ involves the two points $x$ and $x+r$ in the flow field $v(x)$, the statistics of $u(r)$ is a two-point statistics of $v(x)$.

\subsection{Established turbulence models} \label{sec:turbmods}
We now give a brief survey on some general results that have been found for the statistics of the velocity increments $u(r)$ of developed turbulence. An exact result from the Navier-Stokes equation in the limit of infinite Reynolds numbers was found by Kolmogorov in 1941 \cite{Kolmogorov1941c} and states that the third-order structure function depends linearly on $r$,
\begin{equation} \label{eq:four-fifths}
	S^3(r) = -\frac{4}{5}\beps\,r \,.
\end{equation}
This result is known as the \textit{four-fifth law}.

In a series of publications \cite{Kolmogorov1941a,Kolmogorov1941b,Kolmogorov1941c,Obukhov1941a,Obukhov1941b}, Kolmogorov and Obukhov addressed the expected universality of developed turbulence with the conclusion that universality should manifest itself in scaling laws 
\begin{equation} \label{eq:self-sim}
	S^n(r) \simeq S^n(L) \left(\frac{r}{L}\right)^{\z_n} \propto r^{\z_n}
\end{equation}
with universal scaling exponents $\z_n$. From a dimensional analysis it follows that the scaling exponents should be
\begin{equation}\label{eq:zn_K41}
	\z_n = \frac{n}{3} \qquad \text{(K41)}\,, 
\end{equation}
which is now known as K41 scaling. This scaling law was found to hold only for the first few structure functions.

In 1962, Kolmogorov and Obukhov refined the K41 scaling by considering fluctuations of the energy transfer rate $\beps$ \cite{Kolmogorov62JoFM,Oboukhov1962},
\begin{equation}\label{eq:zn_K62}
	\z_n = \frac{n}{3} + \frac{\mu}{18}(3n-n^2) \qquad \text{(K62)} \,.
\end{equation}
The extra term with the intermittency factor $\mu$ is often referred to as intermittency correction, as it accounts for the intermittent fluctuations on small scales not considered in the K41 model. The value of $\mu$ was determined experimentally to be $\mu\approx0.25$ \cite{Arneodo1996}. Since then, many intermittency corrections have been put forward.

A popular intermittency correction was postulated by She and Leveque \cite{SheLeveque94PRL,She1995,Liu2003} by considering a hierarchy of fluctuating structures on dissipative scales. Coarse graining to inertial range scales led to the She-Leveque (SL) scaling law 
\begin{equation} \label{eq:zn_SL_gen}
	\z_n = \left(1-\frac{C_0}{3}\right)\frac{n}{3} + C_0\left(1 - \beta^{\frac{n}{3}}\right) \qquad\text{(SL)} \,.
\end{equation}
Here, $C_0$ is the co-dimension of the dominant fluctuating structure and $1-\beta$ determines the degree of small-scale intermittency in the model: $\beta=1$ corresponds to no intermittency and $\beta=0$ to strongest intermittency. She and Leveque determined from their theory that $\beta=\frac{2}{3}$. For $\beta=1$ and co-dimension $C_0=0$, K41 scaling is recovered. Taking vortex filaments with fractal dimension $\dfr=1$ (i.e. $C_0=3-\dfr=2$) as dominant fluctuating structures and plugging in $\beta=\frac{2}{3}$, the SL scaling becomes parameter-free,
\begin{equation} \label{eq:zn_SL_parfree}
	\z_n = \frac{n}{9} + 2\left[1 - \left(\frac{2}{3}\right)^{\frac{n}{3}}\right] \,.
\end{equation}
The above scaling law is in agreement with all structure functions that can be reliably obtained from measured data \cite{Frisch95}.

The discussed scaling laws only hold for infinite Reynolds numbers or well within the inertial range, as well as for a homogeneous and isotropic flow field. To accommodate for experimental imperfections that do not meet these conditions, Benzi et al. proposed a correction to pure scaling what they call {\it extended self-similarity} (ESS) \cite{Benzi1993,Benzi1996,Chakraborty2010}, which essentially amounts to
\begin{equation} \label{eq:ESS_Sn}
	S^n(r) \propto \big[S^3(r)\big]^{\z_n} \qquad \text{(ESS)} \,.
\end{equation}
The idea is that experimental imperfections and not fully satisfied conditions of developed turbulence have the same impact on all structure functions and can be measured by the deviation of $S^3(r)$ from the four-fifth law (\ref{eq:four-fifths}). In an ideal situation the four-fifth law (\ref{eq:four-fifths}) implies $S^3(r)\propto r$ and usual scaling (\ref{eq:self-sim}) is recovered. In the extensive experimental investigation \cite{Arneodo1996} it has been found that ESS is in excellent agreement with measured data.

Close to the picture of a turbulent cascade are random cascade models (RCMs) \cite{Castaing1995,Castaing1996,Chilla1996,Dubrulle2000,Chevillard2005}. RCMs express $u(r)$ in terms of a multiplier $h(r)$, that is $u(r)=h(r)u(r=L)$, where now $h(r)$ is considered the random variable instead of $u(r)$. On the level of $p(u,r)$, a propagator $G_{rL}(\ln h)$ is used to express $p(u,r)$ as the propagation of $p(u,r=L)$ down to smaller scales $r<L$,
\begin{equation} \label{eq:random-cascade}
	p(u,r) = \int G_{rL}(\ln h)\; p\left(\frac{u}{h},r=L\right) \,\frac{\dd\ln h}{h}  \qquad \text{(RCM)}\,.
\end{equation}
The choice of $G_{rL}(\ln h)$ determines different turbulence models, special cases reproduce K41, K62 and SL scaling. We come back to these special cases when we discuss the underlying Markov process in section \ref{sec:uni}.

The last turbulence model considered here was introduced by Yakhot \cite{Yakhot98PRE} based on a field theoretic approach to Burgers' turbulence \cite{Polyakov95PRE}. The Burgers' equation is basically a simplified Navier-Stokes equation without the pressure term. Yakhot was able to include pressure by using the full Navier-Stokes equation and ended up with a partial differential equation for $p(u,r)$,
\begin{align} \label{eq:yakhot_pde}
	-\frac{\pt\left(u\pt_r\,p(u,r)\right)}{\pt u} + B&\,\frac{\pt p(u,r)}{\pt r} = - \frac{A}{r} \frac{\pt\left(u\,p(u,r)\right)}{\pt u} + \frac{\vrms}{\lch}\frac{\pt^2\left(u\,p(u,r)\right)}{\pt u^2} \,,
\end{align}
with parameters $A$ and $B$. Yakhot determined from his theory that $B\approx20$. Due to the characteristic length scale of turbulence generation $\lch$ and the root-mean-square velocity $\vrms=\sqrt{\lla v^2\rra}$, Yakhot's result includes details of turbulence generation, which is a marking distinction to most other turbulence models.

Integration of the above equation and substitution of $S^n(u,r)=c_nr^{\z_n}$ yields an expression for the scaling exponents,
\begin{align} \label{eq:yakhot_zn_gen}
	\z_n = \frac{An}{B+n} + \frac{r}{\lch}\,\frac{\vrms}{c_n/c_{n-1}}\,\frac{n\,(n-1)}{B+n}\,r^{\z_{n-1}-\z_n} \,.
\end{align}
In the limit of infinite Reynolds numbers, that is $\frac{r}{\lch}\to0$, a pure scaling law remains, 
\begin{align} \label{eq:yakhot_zn_limit}
	\z_n = \frac{An}{B+n} = \frac{n}{3}\frac{B+3}{B+n} \qquad \text{(YAK)} \,,
\end{align}
where $A=\frac{B+3}{3}$ follows from the four-fifth law (\ref{eq:four-fifths}). With Yakhot's prediction $B=20$, also Yakhot's scaling is parameter free. The agreement with experimental data is as good as the SL scaling law. We mention that the above scaling exponent includes Kolmogorov scaling and other scaling laws as lower order terms in a Taylor series \cite{Renner2011}.

\section{Markov approach}\label{sec:markov}
Recall the two properties (a) and (b) of the Navier-Stokes equation (\ref{eq:NSE_dimless}). According to (b), large turbulent structures break up into smaller structures. For a sufficiently large cascade step, the chaotic property (a) then implies that the forming structures only depend on the structures of the previous cascade step. Instead of attempting to resolve the exact dynamics of one cascade step, we take the new structures as a random outcome which only depend on the structures of the previous cascade step. This is essentially the Markov property for the turbulent cascade, in close analogy to Brownian motion.

We begin with demonstrating how such a Markov cascade process can be set up and discuss the implications of the integral fluctuation theorem and second law for this process.

\subsection{Markov cascade process}
The picture of the turbulent cascade from (b) suggests that velocity increments $u(r)$ on scale $r$ are the result of the repeated break-up of the largest structures on scale $L$,
\begin{align} \label{eq:cascade_h}
	u(r)=h_1\cdot\ldots\cdot h_{N(r)}u(L) \,,
\end{align}
where the $h_i$ are the multipliers for each break-up and $N(r)$ is the necessary number of break-ups to reach the scale $r$. According to the chaotic property (a) and if we assume that one cascade step is sufficiently large, the outcome of each break-up is random and only depends on the previous set of structures. We therefore take the $h_i$ to be random numbers and rewrite (\ref{eq:cascade_h}) as
\begin{align} \label{eq:cascade_u}
	\ln \frac{u(r)}{u(L)}=N(r)\ln h_0 + Z(r) \,,
\end{align}
with $h_0$ being the magnitude of the $h_i$ and $Z$ being the sum $Z(r)=\sum_{i=1}^{N(r)}\xi_i$ of new random numbers $\xi_i=\ln(h_i/h_0)$. The value of $h_0$ shall be chosen such that the mean of $Z(r)$ is zero, $\lla Z(r)\rra=0$.

Two assumptions fix the statistical properties of $Z(r)$. Firstly, owing to the homogeneity and isotropy of the flow field, we assume all $\xi_i$ to be drawn from the same distribution. Secondly, due to the chaotic property (a), we assume the $\xi_i$ to be independent, $\lla \xi_i\xi_j\rra=\delta_{ij}$. 

Combining both assumptions, we take the $\xi_i$ to be independent and identically distributed (iid) random numbers. The limit of infinite Reynolds numbers implies infinitely many cascade steps,
\begin{align} \label{eq:cascade_log_u}
	Z(r)=\int_0^{N(r)}\xi(x) \, \dd x \,,
\end{align}
with $\lla\xi(x)\xi(y)\rra=\delta(x-y)$. Applying the central limit theorem we may argue that $Z(r)$ is normal distributed with zero mean and variance $\lla Z(r)^2 \rra=N(r)$. In this continuous limit, (\ref{eq:cascade_log_u}) is the solution of a stochastic differential equation of Langevin type, 
\begin{align} \label{eq:cascade_LE}
 -\frac{\pt}{\pt r} u(r) &= -a\frac{\pt N(r)}{\pt r}\,u(r) + \sqrt{2b\frac{\pt N(r)}{\pt r}\,u(r)^2}\,\xi(r),  \quad u(L) = u_L \,,
\end{align}
where the additional free parameters $a$ and $b$ derive from $\ln h_0$, and we redefined $\xi(r)$ appropriately. The initial value $u_L$ is typically drawn from an initial distribution $p_L(u_L)$. The minus sign is due to evolution in scale from $L$ to smaller scales. In the following, we will refer to this process as the \textit{Markov cascade process}. More details of this process are provided in \cite{DISS}.

Note that a direct test of the above assumptions by numerical or experimental means is intricate since the stochastic force $\xi(r)$ is not readily accessible. However, the Langevin equation (\ref{eq:cascade_LE}) could in principle be solved for $\xi(r)$ and experimentally or numerically determined $u(r)$ be plugged in to analyse the statistical properties of $\xi(r)$.
An attempt along these lines has been made in the experimental studies \cite{MarcqNaert98,MarcqNaert01} confirming the statistical independence of the $\xi_i$, whereas their statistics turned out to be slightly scale dependent indicating that the assumption of identically distributed $\xi_i$ is an approximation. However, this result was obtained for finite Reynolds numbers and a modified Langevin equation has been used.
The extensive experimental study \cite{Renner2002} with Reynolds numbers up to $\Rey\approx10^6$ fills that gap by analysing how the coefficients of the Langevin equation depend on $\Rey$. Indeed, the limiting Langevin equation for $\Rey\to\infty$ was found to be of the form (\ref{eq:cascade_LE}), confirming indirectly the made assumptions.

A similar approach to the cascade process has been discussed in \cite{Benzi1984} in which it is assumed that the cascade takes place in a fractal subspace. The resulting scaling law is known as $\beta$-model \cite{Frisch95}. A superposition of such fractal cascade processes leads to the multifractal model put forward by Frisch and Parisi \cite{FrischParisi1985,Frisch95}. In the limit of infinite many coexisting fractal cascade processes, the application of Laplace's method results in a single scaling law given as the Legendre transformation of the fractal dimensions of the cascades. A Markov process reproducing this scaling law could not be determined, some preliminary results can be found in \cite{DISS} (pages 157--159).

\subsection{Diffusion and jump parts in Markov processes}
Taking the Markov cascade process as a basis, it may serve as a starting ground to develop more realistic turbulence models. By making the connection to the turbulence models introduced in section \ref{sec:turbmods}, we demonstrate in the next section \ref{sec:uni} how such extensions can be made. To do so, it is useful to consider the Fokker-Planck equation for $p(u,r)$ that corresponds to the stochastic differential equation (\ref{eq:cascade_LE}). The Fokker-Planck equation covers the class of continuous Markov processes or diffusion processes. One important step to capture more turbulence models is to extent this class to discontinuous Markov processes or jump processes. In this subsection we recall how to accomplish this extension.

The general Fokker-Planck equation is of the form
\begin{align} \label{eq:FPE}
 -\frac{\pt}{\pt r} p(u,r) = \left[-\frac{\pt}{\pt u}\Df(u,r) + \frac{\pt^2}{\pt u^2}\Dg(u,r)\right]p(u,r) ,\quad p(u,L)=p_L(u)\,.
\end{align}
Interpreting (\ref{eq:cascade_LE}) in the Stratonovich convention, the above Fokker-Planck equation describes the statistics of the Markov cascade process (\ref{eq:cascade_LE}) by the following choice of drift and diffusion coefficients,
\begin{align} \label{eq:cascade_FD}
 \Df(u,r) = -(a-b)\frac{\pt N(r)}{\pt r}\,u ,\quad \Dg(u,r) = b\frac{\pt N(r)}{\pt r}\,u^2 \,.
\end{align}

To add a jump component to (\ref{eq:FPE}), we revert to the Markov property encoded in the Chapman-Kolmogorov relation ($r_1>r_2>r_3$)
\begin{align} \label{eq:CKR}
  p(u_1,r_1|u_3,r_3) = \int p(u_1,r_1|u_2,r_2)\,p(u_2,r_2|u_3,r_3) \, \dd u_2 \,.
\end{align}
By conversion to the differential form of the Chapman-Kolmogorov relation \cite{Gardiner2009},
\begin{align} \label{eq:diffCKR}
  -\frac{\pt p(u,r)}{\pt r}  = &-\frac{\pt}{\pt u} \Df(u,r) p(u,r) + \frac{\pt^2}{\pt u^2} \Dg(u,r) p(u,r) \nonumber\\
  &+ \int \jd(w;u-w,r) p(u-w,r) - \jd(w;u,r) p(u,r) \,\dd w 
\end{align}
with the definitions
\begin{subequations} \label{eq:CKR_3conds}
  \begin{align}
		\label{eq:CKR_3conds_jump}
    &\lim\limits_{\Dr\to0}\,\frac{1}{\Dr}\,p(u,r|u-w,r+\Dr) = \jd(w;u,r) \,, \\ 
    \label{eq:CKR_3conds_drift}
    &\lim\limits_{\Dr\to0}\frac{1}{\Dr} \int\limits_{|w|<\e} w \, p(u,r|u-w,r+\Dr) \,\dd w = \Df(u,r) + \OO(\e) \,, \\
    \label{eq:CKR_3conds_diff}
    &\lim\limits_{\Dr\to0}\frac{1}{2\Dr} \int\limits_{|w|<\e} w^2 \, p(u,r|u-w,r+\Dr) \,\dd w = \Dg(u,r) + \OO(\e) \,,
  \end{align}
\end{subequations}
we find a general evolution equation for Markov processes obeying the Chapman-Kolmogorov relation \cite{Gardiner2009}. Here, we have used the measure $\jd(w;u,r)$ accounting for the probability of a jump from $u$ to $u+w$ at scale $r$. We will refer to $\jd(w;u,r)$ as the jump distribution for the jump width $w$. For $\jd(w;u,r)\equiv0$ we recover the Fokker-Planck equation, while for $\Df(u,r)\equiv\Dg(u,r)\equiv0$ we have a pure jump process governed by the master equation
\begin{equation} \label{eq:ME}
	-\frac{\pt p(u,r)}{\pt r} = \int \jm(u|\tu;r) p(\tu,r) - \jm(\tu|u;r) p(u,r) \, \dd \tu \,,
\end{equation}
where the jump distribution defines the transition probability $\jm(u|\tu,r)$ from $\tu$ to $u$ by $\jm(u|\tu,r)=\jd(u-\tu;\tu,r)$.

It is difficult to map turbulence models to the Master equation directly. A popular indirect approach is to expand $\jd(w;u,r)$ in a power series and work with the Kramers-Moyal expansion of the Master equation \cite{Risken89},
\begin{align} \label{eq:KME}
	-\frac{\pt p(u,r)}{\pt r} = \sum\limits_{k=1}^{\infty}\frac{(-1)^k}{k!}\frac{\pt^k}{\pt u^k}\left[\Ak(u,r)p(u,r)\right] \,,
\end{align}
with the moments of the jump distribution,
\begin{align} \label{eq:def_Ak}
	\Ak(u,r) = \int w^k\jd(w;u,r) \,\dd w \,.
\end{align}
The coefficients $\frac{1}{k!}\Ak(u,r)$ are also known as Kramers-Moyal coefficients. A vanishing even moment implies $\jd(w;u,r)=0$ and as such may serve as an indicator as to whether a measurement series is a realisation of a continuous Markov process, a statement that has been proven by Pawula and is hence known as Pawula's theorem \cite{Pawula1967}.

The definitions (\ref{eq:CKR_3conds_drift}) and (\ref{eq:CKR_3conds_diff}) for the drift and diffusion coefficients are the basis for methods used in Markov analysis to estimate $\Df(u,r)$ and $\Dg(u,r)$ directly from measured turbulence data \cite{FriedrichPeinke97PRL,Renner2001,NawroPeinkKleinFried07PRE,GottschallPeinke08NJoP,Renner2011,Kleinhans2012}. Estimation of the next higher even moments of the jump distribution have been found to be orders of magnitude smaller than the first two moments, implying that the continuous component is indeed the dominant one in the cascade process. However, a truncation of the Kramers-Moyal expansion after the second term can only yield an approximation of the process since small even moments do not imply that all moments are negligibly small. In that context, it should be interesting to estimate the jump distribution by its definition (\ref{eq:CKR_3conds_jump}) directly from measured data.

Note that the Markov analysis of developed turbulence addresses the three-point statistics of the flow field $v(x)$, as it involves velocity increments at two scales, say $r_1$ and $r_2$, which translates into three points in space, $x$, $x+r_1$ and $x+r_2$. In that sense, Markov models of turbulence capture more details than the two-point models introduced in section \ref{sec:turbmods}.

\subsection{Integral fluctuation theorem and second law}
A valuable tool for the estimation and analysis of drift and diffusion coefficients arises from drawing the analogy to stochastic thermodynamics (see, e.g., \cite{Seifert2012} for an overview on stochastic thermodynamics). Small systems exhibit fluctuating heat exchange with their environment, leading to a fluctuating total entropy production $\DS$. Due to the stochastic nature of $\DS$, events with $\DS<0$ are possible, but on average the second law is, of course, still obeyed, $\lla\DS\rra\geq0$. One of the marking results of stochastic thermodynamics is that the second law can be tightened to an equality, the \textit{integral fluctuation theorem} 
\begin{align} \label{eq:iFT}
 \lla e^{-\DS} \rra = 1 \,.
\end{align}
The second law is implied by Jensen's inequality $\langle e^{-x}\rangle\geq e^{-\langle x\rangle}$.

Formally, the total entropy production of a cascade realisation $u(\cdot)$ can be written as \cite{Seifert2005,Nickelsen2013}
\begin{align} \label{eq:DS}
 \DS[u(\cdot)] = \int_L^{r} \frac{\pt u(r')}{\pt r'} \; \frac{\pt}{\pt u}\frac{\Df(u(r'),r')-\frac{\pt}{\pt u}\Dg(u(r'),r')}{\Dg(u(r'),r')} \,\dd r' - \ln\frac{p_r(u_r)}{p_L(u_L)} 
\end{align}
with the initial distribution $p_L(u_L)$ and the solution $p_r(u_r)=p(u(r),r)$ of the Fokker-Planck equation for a smaller scale $r<L$, both are typically obtained from measurements or simulations.

For the Markov cascade process (\ref{eq:cascade_FD}), the integral in the expression for the total entropy production (\ref{eq:DS}) can be solved explicitly and the total entropy production only depends on the initial and final values of the cascade, $u_L$ and $u_r$,
\begin{align} \label{eq:DS_cascade}
 \DS = -\ln\left(\frac{u_r}{u_L}\right)^\nu - \ln\frac{p_r(u_r)}{p_L(u_L)} \,.
\end{align}
The integral fluctuation theorem then reads
\begin{align} \label{eq:cascade_IFT}
 \lla\left(\frac{u_r}{u_L}\right)^\nu\frac{p_r(u_r)}{p_L(u_L)}\rra = 1
\end{align}
implying the second law like inequality
\begin{align} \label{eq:cascade_secondlaw}
	\lla \ln\frac{u_L}{u_r} \rra \geq \frac{1}{\nu}\lla\ln\frac{p_r(u_r)}{p_L(u_L)}\rra 
\end{align}
or
\begin{align} \label{eq:cascade_secondlaw_h}
	\sum\limits_{i=1}^{N(r)} \lla \ln h_i \rra \leq \frac{1}{\nu}\Ds(r) \,,
\end{align}
where we plugged in the multipliers $h_i$ from (\ref{eq:cascade_h}) and denoted the difference in Shannon entropy as $\Ds(r)=-\lla\ln\frac{p_r(u_r)}{p_L(u_L)}\rra$.

We briefly discuss the integral fluctuation theorem (\ref{eq:cascade_IFT}) and second law (\ref{eq:cascade_secondlaw_h}) of the Markov cascade process.
The difference in Shannon entropy $\Ds(r)$ can be shown to be negative for all scales
\cite{DISS}. Hence, the second law (\ref{eq:cascade_secondlaw_h}) states that multipliers must predominantly be smaller than one in order to satisfy the inequality (\ref{eq:cascade_secondlaw}). For $0<h_i<1$, $u(r)$ decreases along the cascade, which is the average tendency of the cascade process and the total entropy production is positive. However, as the second law addresses \textit{averages} of multipliers, rare instances of inverse cascades, $h_i>1$, may occur, resulting into negative values for $\Delta S$. The balance between entropy producing and entropy reducing realisations $u(\cdot)$ has to be such that the integral fluctuation theorem (\ref{eq:iFT}) is satisfied. Due to the exponential average in (\ref{eq:iFT}), a few $u(\cdot)$ with $\DS<0$ outbalance many typical realisations with $\DS>0$.

Although the notion of entropy production is an appealing concept for turbulent cascades, one has to bear in mind that the interpretation of the quantity $\Delta S$ is intricate, as the nature of the conjugate process (inverse cascade) and the source of fluctuations are rather unclear. However, applied to real data, the fluctuation theorem can serve as a sum rule to assess the validity of the Markov process for this data \cite{Nickelsen2013,Reinke2016,Reinke2017}. Furthermore, in \cite{Nickelsen2013} we demonstrated that the fluctuation theorem is in particular sensitive to the correct modelling of small-scale intermittency, as $\DS>0$ arise from strong large-scale fluctuations, whereas the dominant rare realisations with $\DS<0$ exhibit strong small-scale fluctuations together with weak large-scale fluctuations.

For discontinuous Markov processes, the integral fluctuation theorem remains valid, but the entropy production $\DS$ needs to be augmented to account for the entropy change at jumps in $u(\cdot)$. 
This additional entropic contribution to $\DS$ reads \cite{Seifert2012}
\begin{align} \label{eq:entropy_jump}
	S_\mathrm{jump}[u(\cdot)] = \sum\limits_{j=1}^{n} \ln\,\frac{\jm(u^+_j|u^-_j;r_j)}{\jm(u^-|u^+_j;r_j)} \,,
\end{align}
where the jumps are from $u^-_j$ to $u^+_j$ at scales $r_j$, and $n$ is the number of jumps in $u(\cdot)$. Two difficulties complicate the application of an integral fluctuation theorem for $\DS+S_\mathrm{jump}$ to measured turbulence data. Firstly, it is difficult to infer from the experimentally accessible moments $\Ak(u,r)$ the transition probability $\jm(u|\tu,r)$, and secondly, it would be necessary to extract a continuous component from $u(\cdot)$ to be substituted into $\DS[u(\cdot)]$ in (\ref{eq:DS}) in order to plug the remaining jump part into $S_\mathrm{jump}[u(\cdot)]$ above. These are open problems that need to be studied in more detail.

\section{Systematic Markov representations}\label{sec:uni}
In the previous section, we discussed the Markov approach to developed turbulence based on the Markov cascade process. In this section, we build upon this process and demonstrate how systematic modifications lead to the turbulence models discussed in section \ref{sec:turbmods}. In doing so, we partly assemble known results scattered in the literature (K62, RCM, Yakhot), partly present new results (ESS, SL, Yakhot's scaling law) and add new insights on these models on the level of Markov processes. More background on the Markov models can be found in \cite{DISS}, and we mention that \cite{Friedrich2016} similarly addresses the Markov representation of turbulence models with a focus on a numerical study of Burgers' turbulence.

\subsection{Kolmogorov scaling}
We first discuss the Markov cascade process (\ref{eq:cascade_LE}). Integration of the Fokker-Planck equation (\ref{eq:FPE}) with the cascade coefficients (\ref{eq:cascade_FD}) yields a differential equation for the structure functions,
\begin{align} \label{eq:cascade_StrFct}
	-\frac{\pt S^n(r)}{\pt r} = [-(a-b)\,n + b\,n(n-1)] \frac{\pt N(r)}{\pt r} S^n(r) \,.
\end{align}
The solution reads
\begin{align} \label{eq:cascade_StrFct_scaling}
	S^n(r)=S^n(L)\exp\{[an - bn^2]N(r)\} \,.
\end{align}
In the discussion of the Markov cascade process, we left $N(r)$ open. One possibility to specify $N(r)$ is to revert to the cascade picture: The size of a turbulent structure determines the scale $r$ which decreases with each cascade step. Starting with the scale $L$, we can write $r=g_0^{N(r)}L$, where $g_0$ is the average reduction factor for one step. Solving for $N(r)$ then yields $N(r)=\frac{\ln(r/L)}{\ln g_0}$. Plugging this $N(r)$ into (\ref{eq:cascade_StrFct_scaling}) and absorbing $g_0$ into $a$ and $b$ yields the scaling law
\begin{equation}
	\frac{S^n(r)}{S^n(L)}=\left(\frac{r}{L}\right)^{\z_n}, \quad \z_n=an-bn^2 \,.
\end{equation}
This is already the most general scaling law possible for continuous Markov processes. The drift $\Df(u,r)$ determines the linear term in $\z_n$ and the quadratic term is determined by the diffusion coefficient $\Dg(u,r)$. Consequently, only K41 and K62 scaling is covered, which is in agreement with findings in \cite{Hosokawa2002}. For the choice $a=n/3$ and $b=0$, corresponding to a deterministic process with random initial values, we reproduce K41 scaling. In agreement with \cite{FriedrichPeinke97PRL,Renner2001}, we obtain K62 for $a=(2+\mu)/6$ and $b=\mu/18$, that is $\nu=(6+4\mu)/\mu=28$ for $\mu=0.25$ in the fluctuation theorem (\ref{eq:cascade_IFT}) and the second law (\ref{eq:cascade_secondlaw_h}).

The stochastic K62 process can be solved by transformation to logarithmic $r$ and $u$ \cite{Gardiner2009, DISS},
\begin{equation} \label{eq:K62_LE_sol}
	u(r) = u(L)\left(\frac{r}{L}\right)^{a}\exp\left[\sqrt{2b\ln(L/r)}\,Z\right] \,,
\end{equation}
where $Z$ is a normal distributed random variable with zero mean and variance one.

\subsection{Log-normal random cascade model}
In the Markov description, we can embed Kolmogorov scaling and ESS into the class of RCMs (\ref{eq:random-cascade}). From the solution of the Fokker-Planck equation (\ref{eq:FPE}) for the Markov cascade process (\ref{eq:cascade_FD}) \cite{Risken89, DISS},
\begin{equation}
	p(u,r) = \frac{1}{u\sqrt{4\pi b N(r)}}\int p(u,L)\,\exp\left[-\frac{\left(\ln\frac{u}{u_L}+aN(r)\right)^2}{4bN(r)}\right]\dd u_L \,,
\end{equation}
the connection to RCMs becomes apparent by noting that the above solution is of the propagator form (\ref{eq:random-cascade}), where the Green's function of the Fokker-Planck equation is the propagator
\begin{equation} \label{eq:rcm-propagator}
	G_{rL}(\ln h)	 = \frac{1}{\sqrt{4\pi b N(r)}}\,\exp\left[-\frac{\left(\ln h+aN(r)\right)^2}{4bN(r)}\right]
\end{equation}
with $h=u/u_L$. 

The connection between Markov processes and RCMs is not surprising since they rest on the same conceptual footing. The difference is that Markov processes provide a somewhat more microscopic formalism in terms of scale evolution equations. In other words, it are the solutions or realizations of respective Markov processes that reproduce RCMs, as explicated by the above solutions (\ref{eq:K62_LE_sol}) or (\ref{eq:rcm-propagator}) of the Markov cascade process. The more microscopic formulation of Markov processes allows to identify and interpret mechanisms such as deterministic tendencies and source of fluctuations. Furthermore, Markov processes are more general as they capture besides random cascade models also other turbulence models in a unified language which allows a more detailed comparison between these models. Finally, modifications of RCMs on the level of Markov processes are more straight forward and can easily be tested experimentally by means of the integral fluctuation theorem (\ref{eq:cascade_IFT}). In the following, these statements will be elucidated more closely.

The above propagator is a log-normal distribution for $h$ with mean $-aN(r)$ and variance $2bN(r)$ which corresponds to log-normal RCMs, a correspondence that has also been noticed in \cite{Amblard1999}. The K62 model is reproduced by $N(r)=\ln(r/L)$ and is therefore also known as log-normal model. In the K41 limit, $b\to0$, the propagator becomes a $\delta$-distribution.

Choosing a different function for $N(r)$ changes the basis of the scaling law. In particular, ESS scaling (\ref{eq:ESS_Sn}) is just a special case of a RCM for $N(r)=\ln S^3(r)$. Departure from the four-fifth law therefore implies a deformation of the path in scale along which the cascade evolves. As such a deformation would affect all moments of $u(r)$ in the same way, it is a possible explanation why the basic assumption of ESS, namely that imperfections leading to deviations from the four-fifth law affect all structure functions in a similar way, is valid.

The K62 scaling, the ESS scaling and RCMs of the above type all have the same integral fluctuation theorem (\ref{eq:cascade_IFT}) in common. This fluctuation theorem, however, turns out to be not universally fulfilled for measured data \cite{Nickelsen2013,DISS,Reinke2017}: The exponential average either diverges or converges to a value clearly different from $1$, indicating that the two-point statistics of these models miss an essential aspect of the turbulent cascade captured in the three-point statistics of the Markov approach. A more general class of log-normal RCMs may be considered by assuming a distinct $r$-dependency for drift and diffusion, $a(r)$ and $b(r)$, instead of the common $r$-dependency $N(r)$. This Markov process corresponds to RCMs examined by Castaing et al. \cite{Castaing1990,Castaing1996,Dubrulle2000}. But also for this class of RCMs no functional form of $a(r)$ and $b(r)$ could be found such that the corresponding integral fluctuation theorem is fulfilled in experimental tests \cite{Reinke2017,DISS}.

However, in the experimental analysis \cite{Reinke2017}, the log-normal RCM was extended by adding a $u$-independent term $c(r)$ to $\Dg(u,r)$, acting as a constant source of fluctuations. The integral fluctuation theorem resulting from this diffusion process is fulfilled for various flow types and a broad range of Reynolds numbers. In that extension, the deterministic component $a(r)$ and the additive noise term $c(r)$ were found to be reasonably universal, whereas the multiplicative noise term $b(r)$ significantly depends on the kind of turbulence generation and the Reynolds number. However, a limiting process for infinite Reynolds numbers could not be deduced.

\subsection{She-Leveque scaling}
Random cascade models that do not belong to the log-normal class cannot be written as a diffusion process, as they would require non-Gaussian noise in the corresponding stochastic differential equation. In particular, scaling exponents $\zeta_n$ deriving from diffusion processes are limited to be linear and/or quadratic in $n$. By augmenting the diffusion process with a jump process, it is possible to find a Markov representation of models that go beyond Kolmogorov scaling, e.g. the scaling laws found by She and Leveque or Yakhot, as we demonstrate now. 

We keep $\Df$ and $\Dg$ as in the Markov cascade process (\ref{eq:cascade_FD}) and add the following form of the Kramers-Moyal coefficients,
\begin{align} \label{eq:KMC_scallaw}
	\Ak(u,r) = \dk \frac{\pt N(r)}{\pt r} \,u^k \,.
\end{align}
Integration of the differential Chapman-Kolmogorov relation (\ref{eq:diffCKR}) yields the general form
\begin{align} \label{eq:scal-law_Markov_gen}
	S^{n}(r) =S^n(L)\exp \left\{\left[an - bn^2 - \sum\limits_{k=1}^{n}\binom{n}{k}\dk \right] N(r)\right\}
\end{align}
of a scaling law. For $N(r)=\ln(r/L)$ we again obtain a pure scaling law,
\begin{equation} \label{eq:scallaw_Markov_pure}
	\frac{S^n(r)}{S^n(L)}=\left(\frac{r}{L}\right)^{\z_n}, \quad \z_n=an-bn^2-\sum\limits_{k=1}^{n}\binom{n}{k}\dk \,,
\end{equation}
which can be mapped to existing scaling laws. This \textit{Markov scaling law} is the most general scaling law possible for the class of Markov processes having both diffusion and jump parts.

Note that due to the form of $\Df$, $\Dg$ and $\Ak$, the stochastic dynamics does not develop non-zero odd moments. That means, all scaling laws that can be written in this general form do not have a skewness in the cascade process unless the initial odd moments at integral scale $r=L$ are non-zero. The implication is that skewness in the statistics of $u(r)$ is developed during turbulence generation and the turbulent cascade only transports this initial skewness to smaller scales.

Comparison of the above Markov scaling law with SL scaling (\ref{eq:zn_SL_gen}) yields
\begin{equation} \label{eq:SL_F}
	\Df(u,r)=-\frac{1}{3}\left(1-\frac{C_0}{3}\right)\frac{u}{r}
\end{equation}
for the deterministic component, and a jump process defined by the moments of the jump distribution, 
\begin{equation} \label{eq:SL_moms}
	\Ak(u,r) = C_0\left(\beta^{\frac{1}{3}}-1\right)^k\frac{u^k}{r} \,,
\end{equation}
as the stochastic component. In view of the K62 process, the diffusion component is replaced by a jump process.

To obtain the explicit form of the jump distribution, we need to solve the moment problem
\begin{equation}
	\int w^k \jd(w;u,r) \dd w = \frac{C_0}{r}(-\tb u)^k ,\quad \tb=1-\beta^{\frac{1}{3}}\geq0 \,.
\end{equation}
In this case, it is straightforward to determine $\jd(w;u,r)$. We first write down the characteristic function $\varphi(z;u,r)$ of $\jd(w;u,r)$ by writing $\varphi(z;u,r)$ in terms of the moments of $\jd(w;u,r)$,
\begin{align}
	\varphi(z;u,r)=\frac{C_0}{r}\sum \frac{(iz)^k}{k!}(-cu)^k=e^{-izcu} \,.
\end{align}
The jump distribution then follows as the inverse Fourier transformation
\begin{align}
	\jd(w;u,r) = \frac{C_0}{r}\frac{1}{2\pi}\int e^{-izw}e^{-izcu}\,\dd z = \frac{C_0}{r}\,\delta(w+cu)
\end{align}
using the integral representation of the $\delta$-distribution. The unnormalized transition probability then reads
\begin{equation}
	\jm(u|\tu,r)=\frac{C_0}{r} \,\delta(u-\beta^{\frac{1}{3}}\tu) \,.
\end{equation}
Substitution of $\Df(u,r)$ from (\ref{eq:SL_F}), $\Dg\equiv0$ and the above $\jm(u|\tu,r)$ into (\ref{eq:diffCKR}) finally leads to a master equation for $p(u,r)$,
\begin{equation} \label{eq:SL_PDE}
	-r \frac{\pt}{\pt r} p(u,r) = -\frac{1}{3}\left(1-\frac{C_0}{3}\right)\frac{\pt}{\pt u}up(u,r) + C_0\left[\beta^{-\frac{1}{3}}p\left(\beta^{-\frac{1}{3}}u,r\right) - p(u,r)\right] \,,
\end{equation}
which turns out to be some kind of delay partial differential equation reminiscent of a pure death-process with a linear drift.

\begin{figure}
	\includegraphics[width=0.6\textwidth]{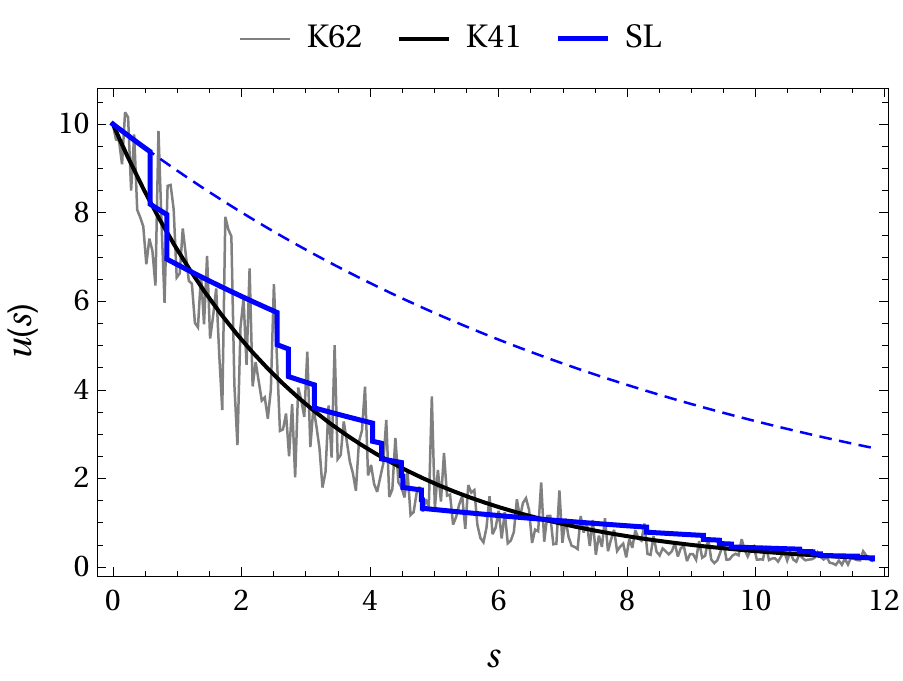}
	\caption{\label{fig:SLprocess} Single realisations of the K41, K62 and SL process on log-scale $s=\ln\frac{L}{r}$ for a fixed initial value $u(s=0)=u(r=L)=10$. The realisation of the K62 process was generated from its solution in (\ref{eq:K62_LE_sol}), the realisation of the SL process was obtained from the simulation procedure explained after (\ref{eq:SL_wait-distr}). The dashed line indicates the deterministic component of the SL process, the black line is the deterministic component of the K62 process, which is the K41 process.}
\end{figure}

To shed some more light on this process, we set up a simulation algorithm that produces realizations $u(\cdot)$ sampling the solution $p(u,r)$ of the above master equation. To do so, we first transform from the scale $r$ to the cascade step $s=\ln\frac{L}{r}$,
\begin{equation}
	\frac{\pt}{\pt s} p(u,s) = -\frac{1}{3}\left(1-\frac{C_0}{3}\right)\frac{\pt}{\pt u}up(u,s) + C_0\left[\beta^{-\frac{1}{3}}p\left(\beta^{-\frac{1}{3}}u,s\right) - p(u,s)\right] \,,
\end{equation}
with the new $p(u,s)=p(u,Le^{-s})$. The escape rate $\gamma$ is then found to be independent from $s$ and $u$,
\begin{equation}
	\gamma=\int \jm(\tu|u,s) \,\dd \tu \equiv C_0 \,,
\end{equation}
with the consequence that the interval $\varDelta$ between jumps is exponentially distributed according to
\begin{equation} \label{eq:SL_wait-distr}
	Q(\varDelta) = C_0 e^{-C_0 \varDelta}
\end{equation}
and independent from the subjacent deterministic process \cite{Gardiner2009}. The simulation procedure hence is to draw $\varDelta$ from the above distribution, let $u(s)$ evolve for this interval according to the linear drift, perform the jump $u\to \beta^{-\frac{1}{3}} u$ and start again. The resulting stochastic process is a jump process with drift, where the jumps occur randomly but with deterministic widths $\beta^{-\frac{1}{3}}u$. We convinced ourselves that the statistics of the $u(\cdot)$ generated by following this procedure indeed exhibit the scaling law by She and Leveque. A typical realisation of this process is depicted in figure \ref{fig:SLprocess} together with a realisation of the K62 process.

From the simulation procedure after (\ref{eq:SL_wait-distr}) and from figure \ref{fig:SLprocess}, we can discuss the nature of the SL process in some more detail. For the theoretical value of $C_0=2$, the deterministic component causes a decrease of $u(r)$ with an exponent of $\frac{1}{9}$, significantly smaller than the value of $\frac{1}{3}$ for the K41 process. However, as the jumps in $u(r)$ are always negative, the decrease of $u(r)$ is comparable for both processes. The fact that $u(r)$ never increases may seem peculiar. But since it is clear from the cascade picture that the deterministic decrease of $u(r)$ goes with $r^\frac{1}{3}$ (like in the K41 process), we could subtract this behaviour from the SL process in order to get the fluctuating part. In the remaining process we would then also find positive fluctuations, as is evident from figure \ref{fig:SLprocess}. In other words, the jump component in the SL process includes a deterministic component.

The intermittency parameter $\beta$ determines the jump widths $w$ in velocity increments: the lower $\beta$, the larger the jumps in $u$, but never larger than $u$ at the instant of the jump, thus, jumps can not overshoot $u=0$. For $\beta=0$ the jump widths are equal to the value of $u$ in the instant the jump occurs, as a consequence, the process remains at the fix-point $u=0$ after the first jump. For $\beta=1$ only the deterministic component remains, which for $C_0=0$ consistently becomes the K41 process. Between these two extreme cases lies the theoretical value of $\beta=\frac{2}{3}$ of the intermittency parameter.

The She-Leveque process is hence a generalisation of the K41 process in terms of adding a jump process to the deterministic K41 process, in contrast to adding continuous diffusion as in the K62 process. 

\subsection{Yakhot}
Comparison of the scaling law (\ref{eq:yakhot_zn_limit}) predicted by Yakhot with the Markov scaling law (\ref{eq:scallaw_Markov_pure}) leads to a pure jump process with the following moments of the jump distribution,
\begin{equation} \label{eq:yakhot_Ak_scal}
	\Ak(u,r) = (B+3)\frac{(-1)^k}{3\tB}\frac{u^k}{r} ,\quad \tB=\prod_{j=1}^k (B+j) \,.
\end{equation}
In this case, the moment problem that would yield the jump distribution could not be solved.

The above jump process emerges as a special case from the Kramers-Moyal expansion
\begin{equation} \label{eq:yakhot_Ak_gen}
	\Ak(u,r) = \left((B+3)-\frac{r/u}{\lch/\vrms}(3B+3)\right)\frac{(-1)^k}{3\tB}\frac{u^k}{r} \,,
\end{equation}
which was found in \cite{Davoudi1999,Davoudi2000} to be equivalent to Yakhot's partial differential equation for $p(u,r)$ in (\ref{eq:yakhot_pde}).

For turn-over times $r/u$ much smaller than $\lch/\vrms$ associated with turbulence generation, that is for very large Reynolds numbers, the extra term in (\ref{eq:yakhot_Ak_gen}) becomes negligible and we recover the scaling law (\ref{eq:yakhot_zn_limit}) implied by the jump process (\ref{eq:yakhot_Ak_scal}). Since the dynamics defined by (\ref{eq:yakhot_Ak_gen}) includes details of turbulence generation and develops a skewness in the statistics of $u(r)$, whereas by the limit $r/u\ll\lch/\vrms$ we omit details of turbulence generation and the process becomes invariant under $u\mapsto-u$, we find again that skewness is developed during turbulence generation and only transported to smaller scales by the turbulent cascade of developed turbulence.

Another special case of (\ref{eq:yakhot_Ak_gen}) follows by noting that the product $\tB$ becomes rapidly smaller with increasing $k$. It therefore is reasonable to only take the first two moments to define an approximate continuous process, $\Df(u,r)=\Af(u,r)$ and $\Dg(u,r)=\frac{1}{2}\Ag(u,r)$, as discussed in \cite{Davoudi1999}. We add that in the limit of infinite Reynolds numbers ($\lch\gg r$), the continuous approximation of Yakhot's model acquires the K62 form and predicts $\mu=\frac{6}{B}=0.3$. This prediction, obtained in the Markov representation, is close to the value $\mu\approx0.25$ found in experiments \cite{Arneodo1996}.

\section{Conclusion}\label{sec:concl}
\begin{table*}
\begin{ruledtabular}
\begin{tabular}{lccccc}
\textrm{Model} & \textrm{Specifics} & $\Df(u,r)$ & $\Dg(u,r)$ & $\Ak(u,r)$ & $\jm(u|\tu,r)$ \\
\colrule
K41 & $\z_n=\frac{n}{3}$ & $-\frac{1}{3}\frac{u}{r}$ & 0 & 0 & 0\\
K62 & $\z_n=\frac{n}{3}+\frac{\mu}{18}\left(3n-n^2\right)$ & $-\frac{3+\mu}{9}\frac{u}{r}$ & $\frac{\mu}{18}\frac{u^2}{r}$ & 0 & 0\\
RCMs & $G_{rL}(\ln h)$ & $-a(r)u$ & $b(r)u^2$ & 0 & 0\\
ESS & $S^n(r)\propto \left[S^3(r)\right]^{\z_n}$ & $-a_0\frac{\pt\ln S^3(r)}{\pt r} u$ & $b_0\frac{\pt \ln S^3(r)}{\pt r} u^2$ & 0 & 0\\
SL & \begin{tabular}{@{}c@{}}$\z_n = \left[1-\frac{C_0}{3}\right]\frac{n}{3}$ \\$\qquad+ C_0\left[1 - \beta^{\frac{n}{3}}\right]$\end{tabular} & $-\frac{1}{3}\left(1-\frac{C_0}{3}\right)\frac{u}{r}$ & 0 & $C_0\left(\beta^{\frac{1}{3}}-1\right)^k\frac{u^k}{r}$ & $\frac{C_0}{r}\delta(u-\beta^{\frac{1}{3}}\tu)$\\
Yakhot & $\dd p(u,r)$ & 0 & 0 & $\left(B_1-B_2(u,r)\right)\psi_k\frac{u^k}{r}$ & ? \\
 & $\z_n=\frac{n}{3}\frac{B+3}{B+n}$ & 0 & 0 & $B_1\psi_k\frac{u^k}{r}$ & ? \\
experimental & $p(u,r|u_L,L)$ & $-a(r)u$ & $b(r)u^2+c(r)$ & 0 & 0
\end{tabular}
\end{ruledtabular}
\caption{\label{tabres} Overview of Markov representations of turbulence models in terms of drift $F(u,r)$, diffusion $D(u,r)$, Kramers-Moyal coefficients $\frac{1}{k!}\Ak(u,r)$ and transition probability $\jm(u|\tu,r)$. The turbulence models are specified by scaling exponents $\z_n$, propagator $G_{rL}$, structure functions $S^n(r)$, partial differential equation $\dd p(u,r)$ or conditional probability $p(u,r|u_L,L)$. The Markov cascade process (\ref{eq:cascade_LE}) is a special case of RCMs for $a(r)=a\frac{\pt N(r)}{\pt r}$ and $b(r)=b\frac{\pt N(r)}{\pt r}$. In the case of Yakhot's model we used the abbreviations $B_1=B+3$, $B_2(u,r)=\frac{r/u}{\lch/\vrms}(3B+3)$ and $\psi_k=\frac{(-1)^k}{3\prod_{j=1}^k (B+j)}$. The experimental model is analysed in \cite{Reinke2017}.}
\end{table*}

We have presented a systematic unification of many turbulence models in the language of Markov processes, as put together in table \ref{tabres}.

The phenomenology of the turbulent cascade motivated a Markov cascade process that turned out to already include Kolmogorov scaling, ESS and a class of RCMs. All obey the same integral fluctuation theorem and second law for the cascade. Although the second law puts a bound on the multipliers of the cascade, we found that it still allows for inverse cascades, which are necessary for the exponential average in the fluctuation theorem to converge to a finite value.

For measured data, the integral fluctuation theorem for the Markov cascade process does not hold universally. Nevertheless, K62 scaling, ESS and RCMs are useful for predicting universal properties of two-point statistics of turbulent flows. For an integral fluctuation theorem to hold for measured data, an additive noise term has to be added to the process, and the multiplicative noise term needs to be tailored to different flow conditions. In other words, a continuous Markov process is not suitable to formulate a universal fluctuation theorem. A promising next step could be to augment diffusion processes with a jump component such that the resulting fluctuation theorem holds universally for measured data.

In this work, we made progress in this regard. Starting from the scaling law for continuous Markov processes, we found that the drift coefficient fixes the term in $\z_n$ that is linear in $n$, and the diffusion coefficient allows for a quadratic term in $\z_n$. Hence, only K41 and K62 scaling laws are covered by diffusion processes. To go beyond Kolmogorov scaling, we derived a Markov scaling law which is the most general form of a scaling law for the class of Markov processes having both diffusion and jump parts. In the Markov description it is clear that every scaling law of this form cannot develop a skewness for the statistics of $u(r)$ but only transports an initial skewness at the integral scale to smaller scales. We demonstrated that the scaling laws found by She and Leveque and by Yakhot are special cases of the Markov scaling law.

For the SL scaling law we were able to derive the jump distribution and set up a master equation. From the master equation we deduced a simulation procedure and discussed the typical realisations of the SL process obtained from this procedure, leading to an interpretation of the parameters of the SL scaling law: The co-dimension $C_0$ is the rate of the exponential distribution which governs the random occurrence of the jumps, the intermittency parameter $\beta$ fixes the change of $u(r)$ at the instances of the jumps. 

Mapping the Markov scaling law to Yakhot's scaling law, we found a pure jump process in terms of Kramers-Moyal coefficients. Determining the jump distribution from the Kramers-Moyal process, however, remains an open problem.

Further open problems are the interpretation of the entropy production based on fundamental equations instead of a mere analogue of a thermodynamic process. The application of fluctuation theorems for Markov processes with jump parts to real data is also an outstanding task, as well as finding the Markov process equivalent to the multifractal model. Also, modifying turbulence models on the level of Markov processes, e.g. adding a diffusion term to the SL process or a jump process to the K62 process, constitutes a novel way of modelling the turbulent cascade of developed turbulence. The conceptual idea of the Markov cascade process may serve as a starting point for this kind of Markov modelling. Finally, it would be interesting to explore the possibilities of synthetically generating flow fields $v(x)$ from solutions or simulations of the Markov models discussed here.

\begin{acknowledgments}
The author thanks Joachim Peinke, Nico Reinke, Andreas Engel and Hugo Touchette for stimulating discussions and acknowledges financial support by the National Institute for Theoretical Physics.
\end{acknowledgments}

\bibliography{turbmods}

\end{document}